\documentclass[two column, prA]{revtex4}%
\usepackage{amsmath}
\usepackage{graphicx}
\usepackage{amsfonts}
\usepackage{amssymb}
\usepackage{color}
\usepackage{bm}%
\providecommand{\U}[1]{\protect\rule{.1in}{.1in}}
\ifx\pdfoutput\relax\let\pdfoutput=\undefined\fi
\newcount\msipdfoutput
\ifx\pdfoutput\undefined\else
\ifcase\pdfoutput\else
\ifx\paperwidth\undefined\else
\ifdim\paperheight=0pt\relax\else\pdfpageheight\paperheight\fi
\ifdim\paperwidth=0pt\relax\else\pdfpagewidth\paperwidth\fi
\fi\fi\fi
\begin{document}
\title{Photon position eigenvectors, Wigner's little group and Berry's phase}
\author{Margaret Hawton}
\email{margaret.hawton@lakeheadu.ca}
\affiliation{Department of Physics, Lakehead University, Thunder Bay, ON, Canada, P7B 5E1}
\author{Vincent Debierre}
\email{debierrev@mpi-hd.mpg.de}
\affiliation{Max Planck Institute for Nuclear Physics, Saupfercheckweg 1, 69117,
Heidelberg, Germany}

\begin{abstract}
We show that the cylindrical symmetry of the eigenvectors of the photon
position operator with commuting components, $\widehat{\mathbf{x}}$, reflects
the $E(2)$ symmetry of the photon little group. The eigenvectors of
$\widehat{\mathbf{x}}$\ form a basis of localized states that have definite
angular momentum, $\widehat{\mathbf{J}}$, parallel to their common axis of
symmetry. This basis is well suited to the description of "twisted light" that
has been the subject of many recent experiments and calculations. Rotation of
the axis of symmetry of this basis results in the observed Berry phase
displacement. We prove that $\left\{  \widehat{x}_{1},\widehat{x}%
_{2},\widehat{J}_{3}\right\}  $ is a realization of the two dimensional
Euclidean $e\left(  2\right)  $ algebra that effects genuine infinitesimal
displacements in configuration space.

\end{abstract}
\maketitle

\section{
\ \ \ \ \ \ \ \ \ \ \ \ \ \ \ \ \ \ \ \ \ \ \ \ \ \ \ \ \ \ \ \ \ \ \ \ \ \ \ \ \ \ \ \ \ \ \ \ \ \ \ \ \ \ \ \ \ \ \ \ \ \ \ \ \ \ \ \ \ \ \ \ \ \ \ \ \ \ \ \ \ \ \ \ \ \ \ \ \ \ \ \ \ \ \ \ \ \ \ \ \ \ \ \ \ \ \ \ \ \ \ \ \ \ \ \ \ \ \ \ \ \ \ \ \ \ \ \ \ \ \ \ \ \ \ \ \ \ \ \ \ \ \ \ \ \ \ \ \ \ \ \ \ \ \ \ \ \ \ \ \ \ \ \ \ \ \ \ \ \ \ \ \ \ \ \ \ \ \ \ \ \ \ \ \ \ \ \ \ \ \ \ \ \ \ \ \ \ \ \ \ \ \ \ \ \ \ \ \ \ \ \ \ \ \ \ \ \ \ \ \ \ \ \ \ \ \ \ \ \ \ \ \ \ \ \ \ \ \ \ \ \ \ \ \ \ \ \ \ \ \ \ \ \ \ \ \ \ \ \ \ \ \ \ \ \ \ \ \ \ \ \ \ \ \ \ \ \ \ \ \ \ \ \ \ \ \ \ \ \ \ \ \ \ \ \ \ \ \ \ \ \ \ \ \ \ \ \ \ \ \ \ \ \ \ \ \ \ \ \ \ \ \ \ \ \ \ \ \ \ \ \ \ \ \ \ \ \ \ \ \ \ \ \ \ \ \ \ \ \ \ \ \ \ \ \ \ \ \ \ \ \ \ \ \ \ \ \ \ \ \ \ \ \ \ \ \ \ \ \ \ \ \ \ \ \ \ \ \ \ \ \ \ \ \ \ \ \ \ \ \ \ \ \ \ \ \ \ \ \ \ \ \ \ \ \ \ \ \ \ \ \ \ \ \ \ \ \ \ \ \ \ \ \ \ \ \ \ \ \ \ \ \ \ \ \ \ \ \ \ \ \ \ \ \ \ \ \ \ \ \ \ \ \ \ \ \ \ \ \ \ \ \ \ \ \ \ \ \ \ \ \ \ \ \ \ \ \ \ \ \ \ \ \ \ \ \ \ \ \ \ \ \ \ \ \ \ \ \ \ \ \ \ \ \ \ \ \ \ \ \ \ \ \ \ \ \ \ \ \ \ \ \ \ \ \ \ \ \ \ \ \ \ \ \ \ \ \ \ \ \ \ \ \ \ \ \ \ \ \ \ \ \ \ \ \ \ \ \ \ \ \ \ \ \ \ \ \ \ \ \ \ \ \ \ \ \ \ \ \ \ \ \ \ \ \ \ \ \ \ \ \ \ \ \ \ \ \ \ \ \ \ \ \ \ \ \ \ \ \ \ \ \ \ \ \ \ \ \ \ \ \ \ \ \ \ \ \ \ \ \ \ \ \ \ \ \ \ \ \ \ \ \ \ \ \ \ \ \ \ \ \ \ \ \ \ \ \ \ \ \ \ \ \ \ \ \ \ \ \ \ \ \ \ \ \ \ \ \ \ \ \ \ \ \ \ \ \ \ \ \ \ \ \ \ \ \ \ \ \ \ \ \ \ \ \ \ \ \ \ \ \ \ \ \ \ \ \ \ \ \ \ \ \ \ \ \ \ \ \ \ \ \ \ \ \ \ \ \ \ \ \ \ \ \ \ \ \ \ \ \ \ \ \ \ \ \ \ \ \ \ \ \ \ \ \ \ \ \ \ \ \ \ \ \ \ \ \ \ \ \ \ \ \ \ \ \ \ \ \ \ \ \ \ \ \ \ \ \ \ \ \ \ \ \ \ \ \ \ \ \ \ \ \ \ \ \ \ \ \ \ \ \ \ \ \ \ \ \ \ \ \ \ \ \ \ \ \ \ \ \ \ \ \ \ \ \ \ \ \ \ \ \ \ \ \ \ \ \ \ \ \ \ \ \ \ \ \ \ \ \ \ \ \ \ \ \ \ \ \ \ \ \ \ \ \ \ \ \ \ \ \ \ \ \ \ \ \ \ \ \ \ \ \ \ \ \ \ \ \ \ \ \ \ \ \ \ \ \ \ \ \ \ \ \ \ \ \ \ \ \ \ \ \ \ \ \ \ \ \ \ \ \ \ \ \ \ \ \ \ \ \ \ \ \ \ \ \ \ \ \ \ \ \ \ \ \ \ \ \ \ \ \ \ \ \ \ \ \ \ \ \ \ \ \ \ \ \ \ \ \ \ \ \ \ \ \ \ \ \ \ \ \ \ \ \ \ \ \ \ \ \ \ \ \ \ \ \ \ \ \ \ \ \ \ \ \ \ \ \ \ \ \ \ \ \ \ \ \ \ \ \ \ \ \ \ \ \ \ \ \ \ \ \ \ \ \ \ \ \ \ \ \ \ \ \ \ \ \ \ \ \ \ \ \ \ \ \ \ \ \ \ \ \ \ \ \ \ \ \ \ \ \ \ \ \ \ \ \ \ \ \ \ \ \ \ \ \ \ \ \ \ \ \ \ \ \ \ \ \ \ \ \ \ \ \ \ \ \ \ \ \ \ \ \ \ \ \ \ \ \ \ \ \ \ \ \ \ \ \ \ \ \ \ \ \ \ \ \ \ \ \ \ \ \ \ \ \ \ \ \ \ \ \ \ \ \ \ \ \ \ \ \ \ \ \ \ \ \ \ \ \ \ \ \ \ \ \ \ \ \ \ \ \ \ \ \ \ \ \ \ \ \ \ \ \ \ \ \ \ \ \ \ \ \ \ \ \ \ \ \ \ \ \ \ \ \ \ \ \ \ \ \ \ \ \ \ \ \ \ \ \ \ \ \ \ \ \ \ \ \ \ \ \ \ \ \ \ \ \ \ \ \ \ \ \ \ \ \ \ \ \ \ \ \ \ \ \ \ \ \ \ \ \ \ \ \ \ \ \ \ \ \ \ \ \ \ \ \ \ \ \ \ \ \ \ \ \ \ \ \ \ \ \ \ \ \ \ \ \ \ \ \ \ \ \ \ \ \ \ \ \ \ \ \ \ \ \ \ \ \ \ \ \ \ \ \ \ \ \ \ \ \ \ \ \ \ \ \ \ \ \ \ \ \ \ \ \ \ \ \ \ \ \ \ \ \ \ \ \ \ \ \ \ \ \ \ \ \ \ \ \ \ \ \ \ \ \ \ \ \ \ \ \ \ \ \ \ \ \ \ \ \ \ \ \ \ \ \ \ \ \ \ \ \ \ \ \ \ \ \ \ \ \ \ \ \ \ \ \ \ \ \ \ \ \ \ \ \ \ \ \ \ \ \ \ \ \ \ \ \ \ \ \ \ \ \ \ \ \ \ \ \ \ \ \ \ \ \ \ \ \ \ \ \ \ \ \ \ \ \ \ \ \ \ \ \ \ \ \ \ \ \ \ \ \ \ \ \ \ \ \ \ \ \ \ \ \ \ \ \ \ \ \ \ \ \ \ \ \ \ \ \ \ \ \ \ \ \ \ \ \ \ \ \ \ \ \ \ \ \ \ \ \ \ \ \ \ \ \ \ \ \ \ \ \ \ \ \ \ \ \ \ \ \ \ \ \ \ \ \ \ \ \ \ \ \ \ \ \ \ \ \ \ \ \ \ \ \ \ \ \ \ \ \ \ \ \ \ \ \ \ \ \ \ \ \ \ \ \ \ \ \ \ \ \ \ \ \ \ \ \ \ \ \ \ \ \ \ \ \ \ \ \ \ \ \ \ \ \ \ \ \ \ \ \ \ \ \ \ \ \ \ \ \ \ \ \ \ \ \ \ \ \ \ \ \ \ \ \ \ \ \ \ \ \ \ \ \ \ \ \ \ \ \ \ \ \ \ \ \ \ \ \ \ \ \ \ \ \ \ \ \ \ \ \ \ \ \ \ \ \ \ \ \ \ \ \ \ \ \ \ \ \ \ \ \ \ \ \ \ \ \ \ \ \ \ \ \ \ \ \ \ \ \ \ \ \ \ \ \ \ \ \ \ \ \ \ \ \ \ \ \ \ \ \ \ \ \ \ \ \ \ \ \ \ \ \ \ \ \ \ \ \ \ \ \ \ \ \ \ \ \ \ \ \ \ \ \ \ \ \ \ \ \ \ \ \ \ \ \ \ \ \ \ \ \ \ \ \ \ \ \ \ \ \ \ \ \ \ \ \ \ \ \ \ \ \ \ \ \ \ \ \ \ \ \ \ \ \ \ \ \ \ \ \ \ \ \ \ \ \ \ \ \ \ \ \ \ \ \ \ \ \ \ \ \ \ \ \ \ \ \ \ \ \ \ \ \ \ \ \ \ \ \ \ \ \ \ \ \ \ \ \ \ \ \ \ \ \ \ \ \ \ \ \ \ \ \ \ \ \ \ \ \ \ \ \ \ \ \ \ \ \ \ \ \ \ \ \ \ \ \ \ \ \ \ \ \ \ \ \ \ \ \ \ \ \ \ \ \ \ \ \ \ \ \ \ \ \ \ \ \ \ \ \ \ \ \ \ \ \ \ \ \ \ \ \ \ \ \ \ \ \ \ \ \ \ \ \ \ \ \ \ \ \ \ \ \ \ \ \ \ \ \ \ \ \ \ \ \ \ \ \ \ \ \ \ \ \ \ \ \ \ \ \ \ \ \ \ \ \ \ \ \ \ \ \ \ \ \ \ \ \ \ \ \ \ \ \ \ \ \ \ \ \ \ \ \ \ \ \ \ \ \ \ \ \ \ \ \ \ \ \ \ \ \ \ \ \ \ \ \ \ \ \ \ \ \ \ \ \ \ \ \ \ \ \ \ \ \ \ \ \ \ \ \ \ \ \ \ Introduction}%

Position operators are key but controversial objects in the study of particle
localization in field theory. Newton and Wigner (NW) found position operators
with commuting components and spherically symmetric eigenvectors for massive
particles and for massless particles with spin $0$ and $\frac{1}{2}$
\cite{NewtonWigner}, but their construction failed for photons. Pryce derived
a photon position operator, $\widehat{\mathbf{x}}_{P}$, consistent with the NW
axioms but its components do not commute so it does not have localized
eigenvectors \cite{Pryce}. A photon position operator with commuting
components, $\widehat{\mathbf{x}}$, does exist \cite{Hawton}, but its
eigenvectors are cylindrically symmetrical and it does not transform like a
vector under rotations \cite{HawtonBaylis}.

We show here that the components of $\widehat{\mathbf{x}}$ perpendicular to
the axis of symmetry of its eigenvectors together with the rotation operator
parallel to this axis are a realization of the two dimensional Euclidean
little group ($E\left(  2\right)  $) \cite{Weinberg}. The operator
$\widehat{\mathbf{x}}$ does not transform like a vector under rotations and
boosts because an additional term is required to rotate the axis of symmetry
of its eigenvectors. This additional term describes the Berry phase
\cite{Berry} displacement of photon position.

The form of the position eigenvectors used here is flexible enough to
accomodate Newton-Wigner and covariant normalization. For photons the
Heisenberg picture position eigenvectors are
\begin{equation}
c_{\sigma\mathbf{x}}^{\mu}(\mathbf{k})=k^{\alpha}e^{\mathrm{i}\left(
\mathbf{k}\cdot\mathbf{x}-kct-\sigma\chi\right)  }\frac{e_{\theta}^{\mu
}+\mathrm{i}\sigma e_{\phi}^{\mu}}{\sqrt{2}} \label{evec}%
\end{equation}
in momentum space spherical polar coordinates where $\sigma$ is helicity,
$\chi$ is the Euler rotation angle about $\mathbf{k}$, and $\mathbf{x}$ is
displacement from the origin. Each choice of $\alpha$ corresponds to a choice
of normalisation for the position eigenvectors. In the NW case for which
$\alpha=1/2$ these eigenvectors are not covariant and they are nonlocal in
configuration space due to the factor $k^{1/2}$. If $\alpha=0$, $c_{\sigma
\mathbf{x}}^{\mu}(\mathbf{k})$ is a four-vector and transformation to
configuration space using the Lorentz invariant measure \textrm{d}$^{3}k/k$ is
a four-vector proportional to the electromagnetic four-potential. The inverse
Fourier transform obtained using the trivial measure is the time derivative of
the vector potential, proportional to the electric field describing this
instantaneouly localized position eigenvector. For the definite helicity
transverse modes $\sigma=$ $\pm$, $\mathbf{B}=-\mathrm{i}\sigma\mathbf{E}$ so
the Riemann-Silberstein vector $\mathbf{E+}\mathrm{i}\sigma\mathbf{B}%
=2\mathbf{E}$ is again an electric field. Details of the normalization of
these position eigenvectors are discussed in \cite{NewtonWigner} and
\cite{HawtonDebierre}, but these details do not affect the expressions derived
in this paper.

The basis of eigenvectors of $\widehat{\mathbf{x}}$ is ideally suited to the
description of optical beams with definite angular momentum (AM) in a fixed
direction. Since orbital AM results in a helical wave front, these beams are
referred to as "twisted light" \cite{Padgett}. It has been observed that total
angular and linear momentum can be transferred from a photon to a particle
trapped in a twisted light beam \cite{ONeil,Zhao}. Focussing of a beam leads
to localization on the axis of symmetry \cite{Zhao} so a basis of localized
states is well suited to the theoretical description of focusing. Berry's
topological phase has been observed using light beams and optical fibers
\cite{Chiao,TomitaChiao,Onoda,Bliokh} as a sideways shift of the beam
centroid. Twisted light beams are currently very topical as they are of
interest as candidates for manipulation of particles, imaging and optical
communications based on violation of local realism \cite{Zeilinger}.

The plan of the paper is as follows: In Section II the Poincar\'{e} and
position operators and their commutation relations are discussed and the AM
and boost operators are separated into intrinsic and extrinsic parts by
writing them in terms of $\widehat{\mathbf{x}}$. In Section III the Wigner
little group algebra is briefly summarized and then extended to include
$\widehat{\mathbf{x}}$ and we prove that the transverse components of
$\widehat{\mathbf{x}}$ together with rotation about its axis of cylindrical
symmetry are a realization of the photon little group. In Section IV
experimental and theoretical work on optical beams is discussed and in Section
V we conclude.

\section{Poincar\'{e} and position operators}

In this Section, after a brief review of the Poincar\'{e} operators, we
introduce $\widehat{\mathbf{x}}$ and its associated Berry phase and then write
the AM and boost operators in terms of position operators.

The Poincar\'{e} group describes the fundamental kinematic symmetry of a
relativistic particle \cite{Stone}. The generators of translations in space
and time, rotations and boosts are the momentum, Hamiltonian, AM and Lorentz
boost operators, $\widehat{\mathbf{P}},$ $\widehat{H},$ $\widehat{\mathbf{J}}
$ and $\widehat{\mathbf{K}}$ respectively. These Poincar\'{e} operators
satisfy the commutation relations $\left[  \widehat{J}_{i},\widehat{J}%
_{j}\right]  =\mathrm{i}\hbar\epsilon_{ijk}\widehat{J}_{k},$ $\left[
\widehat{J}_{i},\widehat{K}_{j}\right]  =\mathrm{i}\hbar\epsilon
_{ijk}\widehat{K}_{k},$ $\left[  \widehat{K}_{i},\widehat{K}_{j}\right]
=-\mathrm{i}\hbar\epsilon_{ijk}\widehat{J}_{k},$ $\left[  \widehat{J}%
_{i},\widehat{P}_{j}\right]  =\mathrm{i}\hbar\epsilon_{ijk}\widehat{P}_{k},$
$\left[  \widehat{K}_{i},\widehat{P}_{j}\right]  =\mathrm{i}\hbar\delta
_{ij}\widehat{H},$ $\left[  \widehat{K}_{i},\widehat{H}\right]  =-\mathrm{i}%
\hbar\widehat{P}_{i},$ $\left[  \widehat{J}_{i},\widehat{H}\right]  =\left[
\widehat{P}_{i},\widehat{H}\right]  =\left[  \widehat{P}_{i},\widehat{P}%
_{j}\right]  =0$ for $i=1,2,3$ \cite{Weinberg}. This algebra will next be
extended to include position operators.

In $\mathbf{k}$-space $\widehat{\mathbf{P}}=\hbar\mathbf{k}$ and the photon
position operator with commuting components, $\widehat{\mathbf{x}}$, is
related to the spinless nonrelativistic momentum space position operator
$\mathrm{i}\mathbf{\partial}_{\mathbf{k}}$ by $\widehat{\mathbf{x}}=k^{\alpha
}\widehat{D}\mathrm{i}\mathbf{\partial}_{\mathbf{k}}\widehat{D}^{-1}%
k^{-\alpha}$ where \cite{HawtonBaylis}
\begin{equation}
\widehat{D}=\exp\left(  -\mathrm{i}\widehat{\sigma}\chi\right)  \exp\left(
-\mathrm{i}\widehat{S}_{3}\phi\right)  \exp\left(  -\mathrm{i}\widehat{S}%
_{2}\theta\right)  . \label{D}%
\end{equation}
Here $\partial_{\mathbf{k}}$ is the $\mathbf{k}$-space gradient,
$\widehat{S}_{i}$ are the Cartesian components of the spin operator
$\widehat{\mathbf{S}}$, $\widehat{\sigma}=\mathbf{e}_{\mathbf{k}}%
\cdot\widehat{\mathbf{S}}$ is the helicity operator, $\theta$ and $\phi$ are
the $\mathbf{k}$-space spherical polar angles, $\chi\left(  \theta
,\phi\right)  $ is the Euler angle and the $\mathbf{k}$-space spherical polar
unit vectors are $\mathbf{e}_{\theta},$ $\mathbf{e}_{\phi}$ and $\mathbf{e}%
_{\mathbf{k}}$ as sketched in Fig. 1. The definite helicity transverse unit
vectors, equal to $\widehat{D}\left(  \mathbf{e}_{1}+\mathrm{i}\sigma
\mathbf{e}_{2}\right)  $, are
\begin{equation}
\mathbf{e}_{\sigma}^{\left(  \chi\right)  }=\frac{1}{\sqrt{2}}\left(
\mathbf{e}_{\theta}+\mathrm{i}\sigma\mathbf{e}_{\phi}\right)  \mathrm{e}%
^{-\mathrm{i}\sigma\chi}. \label{e_chi}%
\end{equation}
The position operator with commuting components is \cite{Hawton,HawtonBaylis}
\begin{equation}
\widehat{\mathbf{x}}=\mathrm{i}\mathbf{\partial}_{\mathbf{k}}-\mathrm{i}%
\alpha\frac{\mathbf{k}}{k^{2}}+\frac{1}{k^{2}}\mathbf{k\times}%
\widehat{\mathbf{S}}-\widehat{\sigma}\mathbf{a}\left(  \theta,\phi\right)
\label{x}%
\end{equation}
where $k=\left\vert \mathbf{k}\right\vert $, $\alpha=\frac{1}{2}$ for the NW
basis and%
\begin{equation}
\mathbf{a}=\frac{\cos\theta}{k\sin\theta}\mathbf{e}_{\phi}+\mathbf{\partial
}_{\mathbf{k}}\chi. \label{a}%
\end{equation}
Inspection of Fig. 1 shows that rotation about $\mathbf{k}$ does not change
$\theta$ or $\phi$. The Euler angle $\chi\left(  \theta,\phi\right)  $ is
defined as a general rotation about $\mathbf{k}$. Any possible transverse
basis is the set of eigenvectors of (\ref{x}) for some $\chi\left(
\theta,\phi\right)  $. Since experiments are often performed on optical beams
with definite angular momentum, the case $\chi=-m\phi$ for which the position
eigenvectors have intrinsic AM $\hbar m\sigma$ in some arbitrary but fixed
direction is of special interest. For this choice of $\chi$ (\ref{a}) becomes
\begin{equation}
\mathbf{a}^{\left(  m\right)  }=\frac{\cos\theta-m}{k\sin\theta}%
\mathbf{e}_{\phi} \label{am}%
\end{equation}

It is known that
\begin{equation}
\widehat{\sigma}a_{i}^{\left(  m\right)  }=\mathrm{i}\mathbf{e}_{\sigma}%
^{\ast}\cdot\mathbf{\partial}_{\mathbf{k}_{i}}\mathbf{e}_{\sigma}
\label{Onoda}%
\end{equation}
is a Berry connection with curvature $\mathbf{\partial}_{\mathbf{k}}%
\times\widehat{\sigma}\mathbf{a}^{\left(  m\right)  }=-\widehat{\sigma
}\mathbf{e}_{k}/k^{2}$ \cite{Onoda,HawtonBaylis,Bliokh3}. For parallel
transport generated by the rotation $\mathrm{d}\bm{\xi}$, we have
$\mathrm{d}\mathbf{k}=\mathrm{d}\bm{\xi}\times\mathbf{k}$, and hence
\begin{equation}
\left(  \mathbf{a}^{\left(  m\right)  }\mathbf{\times k}\right)
\cdot\mathrm{d}\bm{\xi}=-\mathbf{a}^{\left(  m\right)  }\cdot\mathrm{d}%
\mathbf{k,} \label{parallel}%
\end{equation}
so the Berry phase shift is $\sigma\Omega$ where the $m$-independent solid
angle subtended by a loop of the photon's path is \cite{Chiao,HawtonBaylis},%
\begin{equation}
\Omega=-%
{\displaystyle\oint}
\mathbf{a}^{\left(  m\right)  }\cdot\mathrm{d}\mathbf{k}=2\pi\left(
1-\cos\theta\right)  . \label{loop}%
\end{equation}

The position operator with commuting components, $\widehat{\mathbf{x}}$, will
be emphasized here but the properties of the Pryce operator which does not
have commuting components will also be discussed because it is this operator
that is commonly used. The Pryce operator is%
\begin{equation}
\widehat{\mathbf{x}}_{P}=\mathrm{i}\mathbf{\partial}_{\mathbf{k}}%
-\frac{\mathrm{i}}{2}\frac{\mathbf{k}}{k^{2}}+\frac{1}{k^{2}}\mathbf{k\times
}\widehat{\mathbf{S}} \label{Pryce}%
\end{equation}
where, from (\ref{x}),%
\begin{equation}
\widehat{\mathbf{x}}=\widehat{\mathbf{x}}_{P}-\widehat{\sigma}\mathbf{a.}
\label{xP}%
\end{equation}
Since $\left[  \widehat{x}_{Pi},\widehat{x}_{Pj}\right]  =\mathrm{i}%
\epsilon_{ijk}k_{k}/k^{3},$ which can be written as $\widehat{\mathbf{x}}%
_{P}\times\widehat{\mathbf{x}}_{P}=-\mathrm{i}\widehat{\sigma}\mathbf{k/}%
k^{3}$, it is straightforward to verify that the position operator (\ref{x})
does indeed have commuting components: $\widehat{\mathbf{x}}\times
\widehat{\mathbf{x}}=\widehat{\mathbf{x}}_{P}\times\widehat{\mathbf{x}}%
_{P}-\widehat{\sigma}\left(  \mathrm{i}\mathbf{\partial}_{\mathbf{k}%
}\mathbf{\times}\widehat{\mathbf{x}}_{P}\right)  =\mathbf{0}$.

In addition to having commuting components, the position operator (\ref{x})
commutes with the helicity operator and satisfies the usual momentum-position
commutation relations. Using $\widehat{\mathbf{P}}=\hbar\mathbf{k}$ and
$\widehat{H}=\hbar ck$, we write%
\begin{align}
\left[  \widehat{x}_{i},\widehat{x}_{j}\right]   &  =0,\ \left[
\widehat{x}_{i},k_{j}\right]  =\mathrm{i}\delta_{ij},\nonumber\\
\left[  \widehat{x}_{i},k\right]   &  =\mathrm{i}\frac{k_{i}}{k},\ \left[
\widehat{x}_{i},\widehat{\sigma}\right]  =0. \label{commutation}%
\end{align}
In the Heisenberg picture where \textrm{d}$\widehat{O}/\mathrm{d}t=\left[
\widehat{O},\widehat{H}\right]  /\mathrm{i}\hbar$, the momentum space photon
velocity operator is thus%
\begin{equation}
\dot{\widehat{\mathbf{x}}}=c\mathbf{e}_{\mathbf{k}}. \label{v}%
\end{equation}

The Foldy representation \cite{Foldy} of the Poincar\'{e} AM and boost
operators is $\widehat{\mathbf{J}}=\mathrm{i}\hbar\mathbf{\partial
}_{\mathbf{k}}\times\mathbf{k+}\widehat{\mathbf{S}}$ and $\widehat{\mathbf{K}%
}=\frac{1}{2}\mathrm{i}\hbar\left(  k\mathbf{\partial}_{\mathbf{k}%
}+\mathbf{\partial}_{\mathbf{k}}k\right)  +\hbar\mathbf{e}_{\mathbf{k}%
}\mathbf{\times}\widehat{\mathbf{S}}$. In terms of the Pryce position operator
$\widehat{\mathbf{J}}=\hbar\widehat{\mathbf{x}}_{P}\times\mathbf{k}%
+\widehat{\sigma}\hbar\mathbf{e}_{\mathbf{k}},$ $\widehat{\mathbf{K}}=\frac
{1}{2}\hbar\left(  k\widehat{\mathbf{x}}_{P}+\widehat{\mathbf{x}}_{P}k\right)
$. In terms of $\widehat{\mathbf{x}}$ whose eigenvectors, (\ref{e_chi}), are
localized these operators are partitioned into intrinsic and extrinsic parts
\cite{HawtonBaylis}. For the momentum operator%

\begin{align}
\widehat{\mathbf{J}}  &  =\hbar\widehat{\mathbf{x}}\times\mathbf{k+}%
\widehat{\mathbf{J}}^{\left(  0,\mathbf{a}\right)  },\label{J}\\
\widehat{\mathbf{J}}^{\left(  0,\mathbf{a}\right)  }  &  =\widehat{\sigma
}\hbar\left(  \mathbf{a\times k+e}_{\mathbf{k}}\right)  \label{Ja}%
\end{align}
where $\widehat{\mathbf{J}}^{\left(  0,\mathbf{a}\right)  }$ and
$\hbar\widehat{\mathbf{x}}\times\mathbf{k}$ are its intrinsic and extrinsic
parts. The superscript $\left(  0,\mathbf{a}\right)  $ refers to the position
eigenvector at the origin for a particular choice of $\mathbf{a}$. The boost
operator is%
\begin{align}
\widehat{\mathbf{K}}  &  =\frac{\hbar}{2}\left(  k\widehat{\mathbf{x}%
}+\widehat{\mathbf{x}}k\right)  +\widehat{\mathbf{K}}^{\left(  0,\mathbf{a}%
\right)  },\label{K}\\
\widehat{\mathbf{K}}^{\left(  0,\mathbf{a}\right)  }  &  =\widehat{\sigma
}\hbar k\mathbf{a.} \label{Ka}%
\end{align}
Poincar\'{e} transformations are generated by the unitary operator
\cite{Weinberg}
\begin{equation}
\widehat{U}\left(  \bm{\xi},\bm{\beta},\mathbf{x},t\right)  =\exp\left[
\frac{\mathrm{i}}{\mathbf{\hbar}}\left(  \widehat{\mathbf{J}}\cdot
\bm{\xi}\mathbf{-}\widehat{\mathbf{K}}\cdot\bm{\beta}\right)  +\mathrm{i}%
\left(  \widehat{H}t\mathbf{-}\widehat{\mathbf{P}}\cdot\mathbf{x}\right)
/\hbar\right]  \label{U}%
\end{equation}
in which $\bm{\xi}$ is the rotation angle, $\bm{\beta}=\mathbf{v}/c$,
$\mathbf{x}$ a spatial displacement and $t$ is time. The infinitesimal change
in an operator $\widehat{O}$ due to the unitary transformation $\widehat{U}%
^{\dagger}\widehat{O}\widehat{U}$ for $\widehat{U}\left(  \mathrm{d}%
\bm{\xi},\mathrm{d}\bm{\beta},\mathrm{d}\mathbf{x},\mathrm{d}t\right)  $ is
then
\begin{align}
\mathrm{d}\widehat{O}  &  =-\frac{\mathrm{i}}{\hbar}\left\{  \mathrm{d}%
\bm{\xi}\cdot\left[  \widehat{\mathbf{J}},\widehat{O}\right]  -\mathrm{d}%
\bm{\beta}\cdot\left[  \widehat{\mathbf{K}},\widehat{O}\right]  \right\}
\label{Uinf}\\
&  +\frac{\mathrm{i}}{\hbar}\left\{  \mathrm{d}t\left[  \widehat{H}%
,\widehat{O}\right]  -\mathrm{d}\mathbf{x\cdot}\left[  \widehat{\mathbf{P}%
},\widehat{O}\right]  \right\}  .\nonumber
\end{align}

\section{Little group and Wigner translations}

In this Section the properties of the Wigner little group of massless
particles will first be summarized and then their relationship to the position
operators $\widehat{\mathbf{x}}_{P}$ and $\widehat{\mathbf{x}}$ will be
discussed We will prove that the operators $\left\{  \widehat{x}%
_{1},\widehat{x}_{2},\widehat{J}_{3}\right\}  $ are a realization of the
photon little group. Finally, rotation of the axis of symmetry of the basis
will be investigated.

The Wigner little group operators for a specific four-momentum $k^{\mu}$ are
defined by $L_{\nu}^{\mu}k^{\nu}=k^{\mu}$. For a zero mass particle, it is
common, for convenience, to take $\mathbf{k}$ parallel to the $3$-axis,
$k^{\mu}=\left(  k,0,0,k\right)  $, so that the little group operators read
$\widehat{L}=\left\{  \widehat{L}_{1},\widehat{L}_{2},\widehat{J}_{3}\right\}
$, $\widehat{L}_{1}=\widehat{J}_{2}+\widehat{K}_{1}$ and $\widehat{L}%
_{2}=-\widehat{J}_{1}+\widehat{K}_{2}$ \cite{Weinberg,Debierre}. In
$\widehat{L}_{1}$ and $\widehat{L}_{2}$ the component of $\widehat{\mathbf{J}%
}$ needed to compensate for the boost is rotated about $\mathbf{e}_{3}$
relative to $\widehat{\mathbf{K}}$ from $\mathbf{e}_{1}$ to $-\mathbf{e}_{2}$
or from $\mathbf{e}_{2}$ to $\mathbf{e}_{1}$, that is it lags
$\widehat{\mathbf{K}}$ by $\pi/2$. These operators are an $e\left(  2\right)
$ subalgebra of the Poincar\'{e} group that satisfy the commutation relations
$\left[  \widehat{L}_{1},\widehat{L}_{2}\right]  =0,\ \left[  \widehat{J}%
_{3},\widehat{L}_{1}\right]  =\mathrm{i}\hbar\widehat{L}_{2},$ $\left[
\widehat{J}_{3},\widehat{L}_{2}\right]  =-\mathrm{i}\hbar\widehat{L}_{1}$.
Since $\widehat{L}_{1}$ and $\widehat{L}_{2}$ commute they can be
simultaneously diagonalized and their linear combinations have a continuum of
eigenvalues that is not observed, therefore, their common eigenvalue must be
$0$. The operators $\widehat{L}_{1}$ and $\widehat{L}_{2}$ generate gauge
transformations \cite{Weinberg,Kim}.

Photons in a twisted light beam have definite total AM in some fixed direction
that here will be called $\mathbf{e}_{3}$. With $\chi=-m\phi$ the $\mathbf{k}%
$-space unit vectors given by (\ref{e_chi}) become
\begin{equation}
\mathbf{e}_{\sigma}^{\left(  m\right)  }\left(  \mathbf{k}\right)  =\frac
{1}{\sqrt{2}}\left(  \mathbf{e}_{\theta}+\mathrm{i}\sigma\mathbf{e}_{\phi
}\right)  e^{\mathrm{i}\sigma m\phi}. \label{e1}%
\end{equation}
Position eigenvectors at arbitrary $\mathbf{x}$ and $t$ can be obtained by
applying the unitary transformation $U^{\dag}$ given by (\ref{U}) to the
position eigenvectors at the origin, $k^{\alpha}\mathbf{e}_{\sigma}^{\left(
m\right)  }\left(  \mathbf{k}\right)  $, to obtain%
\begin{equation}
\mathbf{c}_{\sigma\mathbf{x}}^{\left(  m\right)  }\left(  \mathbf{k}\right)
=k^{\alpha}\mathbf{e}_{\sigma}^{\left(  m\right)  }\left(  \mathbf{k}\right)
\exp\left[  \mathrm{i}\left(  \mathbf{k}\cdot\mathbf{x}-\omega t\right)
\right]  \label{ex}%
\end{equation}
consistent with $c_{\sigma\mathbf{x}}^{\mu}=\left(  c_{\sigma\mathbf{x}}%
^{0},\mathbf{c}_{\sigma\mathbf{x}}\right)  $ in (\ref{evec}). Experiments are
usually performed on optical beams for which focusing leads to photon
localization in two dimensions. In these beams photon density is independent
of $x_{3}$ and $t$. Using $\int_{-\infty}^{\infty}dz\exp\left[  \mathrm{i}%
\left(  k_{z}-k_{z_{0}}\right)  z\right]  /2\pi=\delta\left(  k_{z}-k_{z_{0}%
}\right)  $ the integral of (\ref{e1}) over $z$ gives
\begin{equation}
\mathbf{e}_{\sigma\bot}^{\left(  m\right)  }\left(  \mathbf{k}_{\bot}\right)
=\mathbf{e}_{\sigma}^{\left(  m\right)  }\left(  \mathbf{k}_{\bot},k_{z_{0}%
}\right)  \exp\left[  \mathrm{i}\left(  \mathbf{k}_{\bot}\cdot\mathbf{x}%
_{\bot}-k_{z_{0}}z-\omega t\right)  \right]  \label{eperp}%
\end{equation}
with $\omega=\sqrt{\mathbf{k}_{\bot}^{2}+k_{z_{0}}^{2}}$. This is a good basis
for the description of optical beams.

To describe rotations about fixed axes, $\mathbf{a}^{\left(  m\right)  }$ will
be written in terms of the Cartesian unit vectors $\mathbf{e}_{1}$,
$\mathbf{e}_{2}$ and $\mathbf{e}_{3}$ sketched in Fig. 1.%
\begin{figure}[ptb]%
\centering
\includegraphics[
natheight=3.261200in,
natwidth=2.355700in,
height=3.2612in,
width=2.3557in
]%
{C:/Users/user/Dropbox/graphics/Fig1__1.pdf}%
\caption{Spherical, cylindrical and Cartesian cordinates}%
\end{figure}
In $\mathbf{k}$-space
\begin{equation}
\mathbf{e}_{\mathbf{k}}=\cos\theta\mathbf{e}_{3}+\sin\theta\mathbf{e}_{\kappa
},\ \mathbf{e}_{\kappa}=\cos\phi\mathbf{e}_{1}+\sin\phi\mathbf{e}%
_{2}.\label{cylindrical}%
\end{equation}
Substitution of (\ref{cylindrical}) in (\ref{Ja}) gives the $\mathbf{e}_{3}$
component of the intrinsic part of the AM operator as%
\begin{equation}
\widehat{J}_{3}^{\left(  0,-m\phi\right)  }=\widehat{\sigma}m\mathbf{\hbar
.}\label{J3}%
\end{equation}
Thus the localized states have intrinsic AM $\hbar\widehat{\sigma}%
m\mathbf{e}_{3}.$

While the momentum, angular momentum and boost operators transform like three
dimensional vectors, for $\widehat{\mathbf{x}}$ operator algebra based on
(\ref{commutation}) and (\ref{J}) to (\ref{Ka}) gives
\begin{align}
\left[  \widehat{J}_{i},\widehat{x}_{j}\right]   &  =\mathrm{i}\hbar
\epsilon_{ijk}\widehat{x}_{k}-\mathrm{i}\partial_{k_{j}}\widehat{J}%
_{i}^{\left(  0,\mathbf{a}\right)  },\label{Jx}\\
\left[  \widehat{K}_{i},\widehat{x}_{j}\right]   &  =-\mathrm{i}\frac{\hbar
}{2}\left(  \frac{k_{j}}{k}\widehat{x}_{i}+\widehat{x}_{i}\frac{k_{j}}%
{k}\right)  -\mathrm{i}\partial_{k_{j}}\widehat{K}_{i}^{\left(  0,\mathbf{a}%
\right)  }. \label{Kx}%
\end{align}
Since $\widehat{J}_{3}^{\left(  0,-m\phi\right)  }=\widehat{\sigma}m\hbar$
given by (\ref{J3}) does not depend on $\mathbf{k}$ in (\ref{Jx}) and the
components of $\widehat{\mathbf{x}}$ commute, in the basis $\chi=-m\phi$
\begin{align}
\left[  \widehat{x}_{1},\widehat{x}_{2}\right]   &  =0,\label{xx}\\
\left[  \widehat{J}_{3},\widehat{x}_{1}\right]   &  =\mathrm{i}\hbar
\widehat{x}_{2},\label{Jx1}\\
\left[  \widehat{J}_{3},\widehat{x}_{2}\right]   &  =-\mathrm{i}%
\hbar\widehat{x}_{1}. \label{Jx2}%
\end{align}
Thus $\left\{  \widehat{x}_{1},\widehat{x}_{2},\widehat{J}_{3}\right\}  $ is a
realization of the two dimensional Euclidean $e\left(  2\right)  $ algebra
that effects genuine infinitesimal transformations in configuration space.
\emph{This is the primary result of this paper.}

The Poincar\'{e}, little group and Pryce position operators are discussed in
\cite{Stone}. The commutators%
\begin{align}
\left[  \widehat{J}_{i},\widehat{x}_{Pj}\right]   &  =\mathrm{i}\hbar
\epsilon_{ijk}\widehat{x}_{Pk},\label{JxP}\\
\left[  \widehat{K}_{i},\widehat{x}_{Pj}\right]   &  =-\mathrm{i}\frac{\hbar
}{2}\left(  \frac{k_{j}}{k}\widehat{x}_{Pi}+\widehat{x}_{Pi}\frac{k_{j}}%
{k}\right)  -\mathrm{i}\widehat{\sigma}\hbar\epsilon_{ijk}\frac{k_{k}}{k^{2}}.
\label{KxP}%
\end{align}
are equivalent to (4) and (10) in \cite{Stone}. The position operator
$\widehat{\mathbf{x}}$ whose components commute has the additional features
that it has an axis of symmetry and localized eigenvectors.

To simplify (\ref{Jx}) and (\ref{Kx}) and obtain their physical interpretation
we will write them in vector form and assume constant $\mathrm{d}\bm{\xi}$,
$\mathrm{d}\bm{\beta}$, \textrm{d}$t$ and \textrm{d}$\mathbf{x}=0$. We return
to the general case since $\mathbf{a}\left(  \theta,\phi\right)  $ is needed
to describe a basis with axis of symmetry not parallel to $\mathbf{e}_{3}$. A
nonzero commutator between $\widehat{\mathbf{x}}$ and $\widehat{\mathbf{J}}$,
$\widehat{\mathbf{K}}$ or $\widehat{H}$ implies a infinitesimal change in the
position operator. For a rotation through the angle \textrm{d}$\bm{\xi}$ and a
velocity change $c$\textrm{d}$\bm{\beta},$ (\ref{Uinf}) gives $\mathrm{d}%
\widehat{x}_{j}^{\xi}=-\left(  \mathrm{i}/\hbar\right)  \mathrm{d}%
\bm{\xi}\cdot\left[  \widehat{\mathbf{J}},\widehat{x}_{j}\right]  $ and
$\mathrm{d}\widehat{x}_{j}^{\beta}=\left(  \mathrm{i}/\hbar\right)
\mathrm{d}\bm{\beta}\cdot\left[  \widehat{\mathbf{K}},\widehat{x}_{j}\right]
$. The corresponding changes in the position operator are%
\begin{align}
\mathrm{d}\widehat{\mathbf{x}}^{\xi}  &  =\mathrm{d}\bm{\xi}\mathbf{\times
}\widehat{\mathbf{x}}-\partial_{\mathbf{k}}\left(  \mathrm{d}\bm{\xi}\cdot
\widehat{\mathbf{J}}^{\left(  0,\mathbf{a}\right)  }\right)  ,\label{dxeta}\\
\mathrm{d}\widehat{\mathbf{x}}^{\beta}  &  =\frac{\mathbf{k}}{k}%
\mathrm{d}\bm{\beta}\cdot\widehat{\mathbf{x}}-\partial_{\mathbf{k}}\left(
-\mathrm{d}\bm{\beta}\cdot\widehat{\mathbf{K}}^{\left(  0,\mathbf{a}\right)
}\right)  . \label{dxbeta}%
\end{align}
The first term on the right hand side of (\ref{dxbeta}) arises because the
energy and position operators do not commute. This corresponds to the
$\mathrm{d}t$ term of (\ref{Uinf}) and is a feature of the quantum mechanics
of both massive and massless particles. The terms in round brackets are
$\widehat{\sigma}$ multiplied by a change in the Euler angle $\chi\left(
\theta,\phi\right)  $ where, in (\ref{dxeta}),
\begin{equation}
\widehat{\sigma}\mathrm{d}\chi^{\xi}=\frac{1}{\hbar}\mathrm{d}\bm{\xi}\cdot
\widehat{\mathbf{J}}^{\left(  0,\mathbf{a}\right)  }\left(  \theta
,\phi\right)  . \label{delta}%
\end{equation}
Rotation about an axis in the $12$-plane will change the axis of symmetry of
the basis. For rotation though an angle \textrm{d}$\theta$ about a fixed axis
that makes an angle $\varphi$ with the $\mathbf{e}_{1}$ axis, $\mathrm{d}%
\bm{\xi}=-\mathrm{d}\theta\left(  \cos\varphi\mathbf{e}_{1}+\sin
\varphi\mathbf{e}_{2}\right)  =-$\textrm{d}$\theta\mathbf{e}_{\mathbf{\kappa
}_{\varphi}}$. For a boost described by $\mathrm{d}\bm{\beta}=\mathrm{d}%
\theta\left(  -\mathrm{\sin}\varphi\mathbf{e}_{1}+\cos\varphi\mathbf{e}%
_{2}\right)  =\mathrm{d}\theta\mathbf{e}_{\varphi}$ leads $\mathrm{d}\bm{\xi}$
by $\pi/2$ so that $\mathrm{d}\chi^{\xi}=-\mathrm{d}\chi^{\beta}=ka^{\left(
m\right)  }\left(  \theta\right)  \cos\left(  \phi-\varphi\right)  $ and the
change in Euler angle introduced by the boost cancels that due to the
rotation. For a finite rotation about $\mathbf{e}_{\mathbf{\kappa}_{\varphi}}$
this Euler angle can be integrated over $\theta$ to give $\Delta\chi^{\xi
}=\left[  \int_{\theta_{i}}^{\theta_{f}}a^{\left(  m\right)  }\left(
\theta\right)  \mathrm{d}\theta\right]  \cos\left(  \phi-\varphi\right)  $.
Since, according to (\ref{x}) and (\ref{a}), $\widehat{\mathbf{x}}$ includes a
term $-\widehat{\sigma}\partial_{\mathbf{k}}\chi\left(  \theta,\phi\right)  $
it follows that
\begin{align}
\mathrm{d}\widehat{\mathbf{x}}^{\xi}-\mathrm{d}\bm{\xi}\times
\widehat{\mathbf{x}}  &  =\mathrm{d}\widehat{\mathbf{x}}^{\beta}%
-\mathbf{e}_{\mathbf{k}}\mathrm{d}\bm{\beta}\cdot\widehat{\mathbf{x}%
}\nonumber\\
&  =-\widehat{\sigma}\mathrm{d}\theta\left[  \mathbf{e}_{\theta}\frac{\partial
a^{\left(  m\right)  }\left(  \theta\right)  }{\partial\theta}\cos\left(
\phi-\varphi\right)  \right. \nonumber\\
&  \left.  +\mathbf{e}_{\phi}\frac{a^{\left(  m\right)  }\left(
\theta\right)  }{\sin\theta}\sin\left(  \phi-\varphi\right)  \right]  .
\label{dx}%
\end{align}

The position operator $\widehat{\mathbf{x}}$ describes the center of AM, while
the Pryce operator $\widehat{\mathbf{x}}_{P}=\widehat{\mathbf{x}%
}+\widehat{\sigma}\mathbf{a}^{\left(  1\right)  }$ implies orbital AM relative
to this center. When applied to the position eigenvector at the origin,
$\widehat{\mathbf{x}}_{P}\mathbf{e}_{\sigma}^{\left(  m\right)  }=\left(
\widehat{\mathbf{x}}+\widehat{\sigma}\mathbf{a}^{\left(  m\right)  }\right)
\mathbf{e}_{\sigma}^{\left(  m\right)  }=\sigma\mathbf{a}^{\left(  m\right)
}\mathbf{e}_{\sigma}^{\left(  m\right)  }$ so the orbital AM is $\sigma
\hbar\mathbf{a}^{\left(  m\right)  }\times\mathbf{k}$. The position operator
$\widehat{\mathbf{x}}$ obeys the commutation relations (\ref{Jx}) and
(\ref{Kx}), while for $\widehat{\mathbf{x}}_{P}$ (\ref{JxP}) and (\ref{KxP})
are satisfied. These commutation relations are similar except that (\ref{Jx})
and (\ref{Kx}) contain a term that rotates the axis of symmetry, while
(\ref{KxP}) contains a term $-\mathrm{i}\widehat{\sigma}\hbar\epsilon
_{ijk}k_{k}/k^{2}$ not present in (\ref{Kx}) due to noncommutativity of the
components of the Pryce position operator. The extra term in (\ref{KxP}) is
equivalent to the second term on the right hand side of (10) and the right
hand side of (13) in \cite{Stone}. Since from (\ref{xP}) and the
$\widehat{\mathbf{x}}_{P}$ commutation relation following it $\left[
\widehat{x}_{i},\widehat{x}_{j}\right]  =\left[  \widehat{x}_{Pi}%
,\widehat{x}_{Pj}\right]  -\widehat{\sigma}\left(  \left[  \widehat{x}%
_{i},a_{j}^{\left(  m\right)  }\right]  +\left[  a_{i}^{\left(  m\right)
},\widehat{x}_{j}\right]  \right)  =\mathrm{i}\widehat{\sigma}\mathbf{\partial
}_{\mathbf{k}}\mathbf{\times}a^{\left(  m\right)  }-\mathrm{i}\widehat{\sigma
}\mathbf{k}/k^{3}=0$, in (\ref{Kx}) this `Wigner' term is absorbed into
$\widehat{\mathbf{x}}$ that has commuting components.

\section{Optical beams}

In this Section application of the position eigenvectors to optical beams will
be discussed in the context of the recent experimental and theoretical
literature. We will consider the relationship of the configuration space basis
to transfer of linear and angular momentum to a particle, focusing, phase
shift in an optical fiber and optical communications.

The linear and angular momentum of a photon can be transferred to a particle.
It is observed that the optical intensity in a high-order Bessel beam is
independent of $t$ and $x_{3}$ and its transverse profile is a series of
bright rings. A small particle trapped in a bright ring of such a beam
simultaneously spins on its axis and orbits the beam centroid \cite{ONeil}. A
photon in this beam has transverse wave vector $\mathbf{k}_{\perp}=k_{\perp
}\mathbf{e}_{\phi},$ a radial position vector $\mathbf{x}_{\perp}$ pointing
outward from the beam axis and extrinsic orbital AM $l\hbar\mathbf{e}_{3}$
\cite{Padgett}. If it is absorbed, its linear momentum $\hbar\mathbf{k}%
_{\perp}$ and total AM $\left(  \sigma+l\right)  \hbar\mathbf{e}_{3}$ will be
transferred to the particle causing it to spin on its axis and orbit the beam
axis \cite{Zhao}. At a fundamental level it is total angular and linear
momentum that is conserved \cite{CT}.

The eigenvectors of $\widehat{\mathbf{x}}$ are an idealization of an
ultrashort pulse focused at $\mathbf{x}$ for an instant. Experiments are
usually performed on optical beams for which focusing leads to photon
localization in only two dimensions. Eq. (\ref{eperp}) is a good basis for
description of focusing of a beam. A CP beam with incident center wave vector
$\mathbf{k}_{i}=k_{i}\mathbf{e}_{3}$ and final wave vector $\mathbf{k}_{f}$ is
focused to the point $\left(  0,0,x_{3}\right)  $ as sketched in Fig. 2.%
\begin{figure}[ptb]%
\centering
\includegraphics[
natheight=2.833100in,
natwidth=2.273600in,
height=2.8331in,
width=2.2736in
]%
{C:/Users/user/Dropbox/graphics/Fig2__2.pdf}%
\caption{Focusing of a light beam.}%
\end{figure}
Since refraction by the lens conserves the component of AM parallel to its
axis of symmetry, the total AM per photon at the focal point is still $\sigma
m\hbar\mathbf{e}_{3}$. This conversion has been observed: focusing of a beam
carrying spin AM can induce orbital AM which drives the orbital motion of
micron-sized metal particles \cite{Zhao}.

In an optical fiber photon position is limited by the diameter of the fiber
and the longitudinal component of momentum is determined by its orientation. A
right-handed CP beam cycling around a closed circuit in $\mathbf{k}$-space
acquires a Berry phase shift relative to a left-handed CP beam of
$2\Omega=4\pi\left(  1-\cos\theta\right)  $ per loop\ as predicted in
\cite{Chiao}, confirmed experimentally in \cite{TomitaChiao}, and given here
by (\ref{loop}). It was predicted in \cite{vanEnk} that electrons and photons
experience a universal geometric phase shift even in a straight waveguide that
can be described in perturbation theory by spin-orbit coupling which in the
paraxial limit is proportional to the spin-orbit coupling. This effect has
recently been observed in dispersion-taylored straight few-mode fibers
\cite{Raymer} where $x_{3}$ was varied by cutting the fiber. The linear
polarization was found to rotate with $x_{3}$ at a rate proportional to the
spin-orbit coupling strength. In these experiments, photons in the input beam
have extrinsic orbital AM $\hbar l_{3}$.

Twisted photons can be used to encode information beyond one bit per single
photon \cite{Zeilinger}. Secure communication requires entangled photons and
entanglement is the most mysterious property of quantum particles. Tests of
quantum mechanics are often performed on photons so the controversy regarding
photon localization is relevant to many experiments that have great potential
for the performance of quantum tasks.

\section{Conclusion}

Photons are the most important but also the most problematic neutral bosons.
They are important because they are the subject of many experiments, including
some intended as tests of quantum mechanics itself. They are controversial
because most theorists believe that there is no acceptable photon position
operator with commuting components that would lead to a basis of position
eigenvectors. But such an operator does in fact exist and we show here that
its properties are a consequence of the symmetry of the photon little group.
The extra term in its commutation relations with the rotation and boost
operators describes rotation of the axis of symmetry of its eigenvectors and
the observed Berry phase shift.

We have proved in Section III that $\left\{  \widehat{x}_{1},\widehat{x}%
_{2},\widehat{J}_{3}\right\}  $ is a realization of the two dimensional
Euclidean $e\left(  2\right)  $ algebra that effects genuine infinitesimal
transformations in configuration space. This answers the question "What is
$\mathbf{x}$" posed by Stone, Dwivede and Zhou \cite{Stone}: $\mathbf{x}$ is
photon position. While still controversial this conclusion illuminates the
debate surrounding photon wave mechanics and the localization of light.

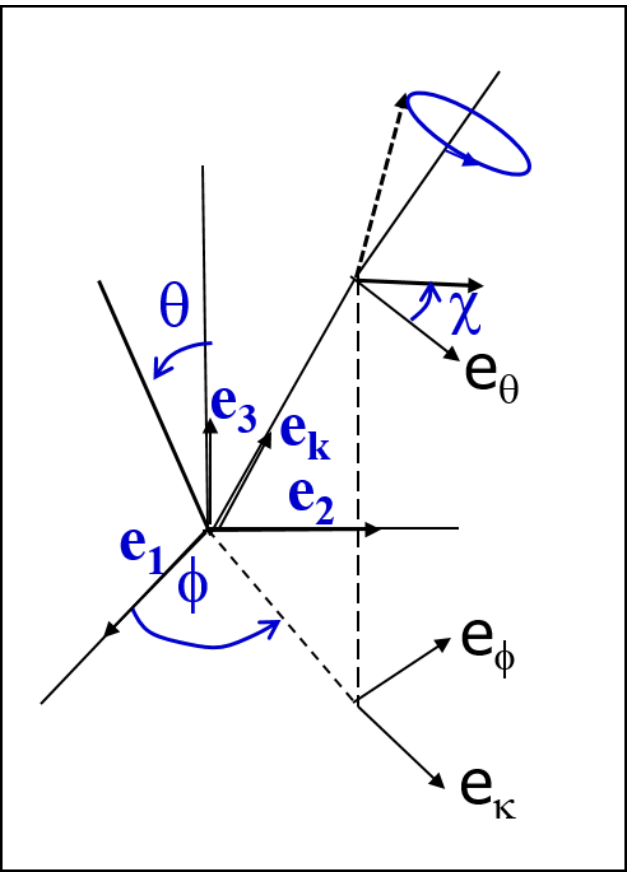
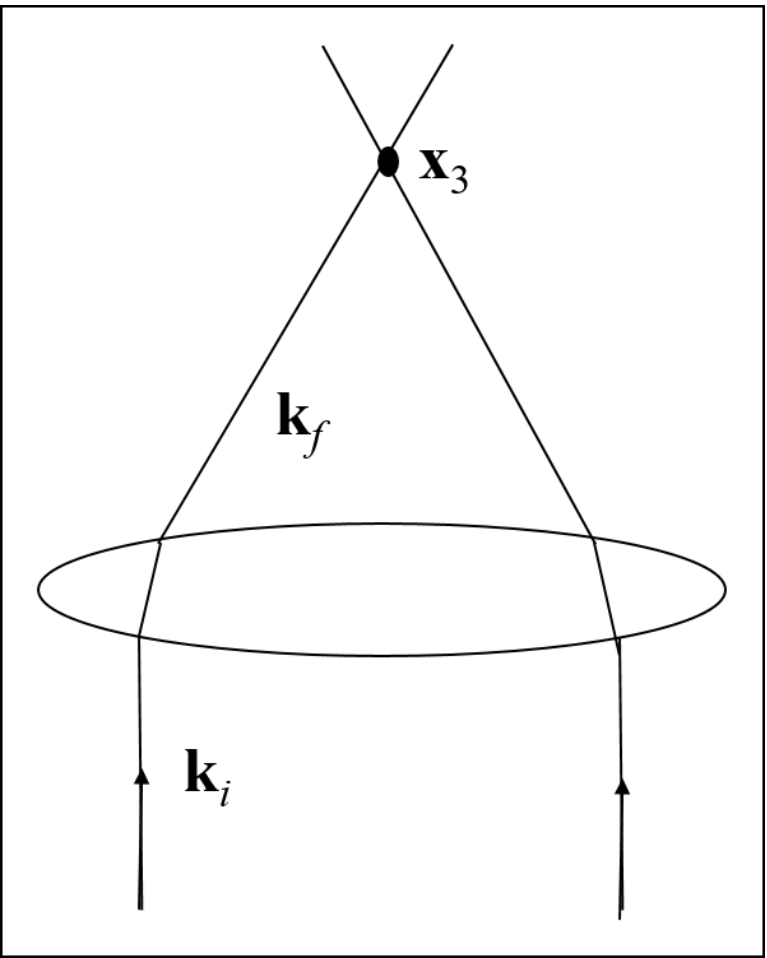
\end{document}